\documentclass[aps,prl,reprint,superscriptaddress,bibnotes,eprint]{revtex4-2}
\usepackage{graphicx}
\usepackage{amsmath}
\usepackage{bbm}
\usepackage{hyperref}
\usepackage{xcolor}
\usepackage{verbatim}
\usepackage{bbold}
\usepackage{float}
\begin{document}
	
	\title{Tensor Networks Can Resolve Fermi Surfaces}
	
	\author{Quinten Mortier}
	\email{quinten.mortier@ugent.be}
	\affiliation{Department of Physics and Astronomy, Ghent University, Krijgslaan 281, S9, 9000 Gent, Belgium}
	
	\author{Norbert Schuch}
	
	\affiliation{Max-Planck-Institute of Quantum Optics,
		Hans-Kopfermann-Stra\ss{}e 1, 85748 Garching, Germany, and\\
		Munich Center for Quantum Science and Technology,
		Schellingstra\ss{}e~4, 80799 M\"unchen, Germany}
	\affiliation{University of Vienna, Faculty of Physics, Boltzmanngasse
		5, 1090 Wien, Austria, and\\
		University of Vienna, Faculty of Mathematics, Oskar-Morgenstern-Platz 1, 1090 Wien, Austria}
	\author{Frank Verstraete}
	\affiliation{Department of Physics and Astronomy, Ghent University, Krijgslaan 281, S9, 9000 Gent, Belgium}
	
	\author{Jutho Haegeman}
	\affiliation{Department of Physics and Astronomy, Ghent University, Krijgslaan 281, S9, 9000 Gent, Belgium}
	
	\begin{abstract}
		
		We demonstrate that projected entangled-pair states (PEPS) are able to represent ground states of critical, fermionic systems exhibiting both 1d and 0d Fermi surfaces on a 2D lattice with an efficient scaling of the bond dimension. Extrapolating finite size results for the Gaussian restriction of fermionic projected entangled-pair states to the thermodynamic limit, the energy precision as a function of the bond dimension is found to improve as a power law, illustrating that an arbitrary precision can be obtained by increasing the bond dimension in a controlled manner. In this process, boundary conditions and system sizes have to be chosen carefully so that nonanalyticities of the \textit{Ansatz}, rooted in its nontrivial topology, are avoided.
		
	\end{abstract}
	
	\maketitle
	
	In one spatial dimension, physically relevant states of quantum many-body systems with a local and gapped Hamiltonian can be represented efficiently by matrix product states (MPS) \cite{MPSorig,reviewTNS}. A natural extension of this construction to higher dimensions was formulated in the form of projected entangled-pair states (PEPS) \cite{PEPSorig,PEPSorigorig}. Both \textit{Ans\"atze} owe their versatile applicability to an inherent area law of entanglement \cite{arealaworig,arealaw}. However, critical systems, with correlations following a power-law decay, can violate this area law in a logarithmic manner. Do tensor networks like MPS and PEPS then still represent efficient \textit{Ans\"atze} for the relevant states of such critical models? In one dimension, this question was already answered in an affirmative way. For gapped and critical models alike, finite-size ground states can be represented faithfully as MPS with a cost that scales polynomially in the system size \cite{1dcrit}. In the thermodynamic limit, local quantities can still be obtained efficiently from MPS ground state approximations, even for critical systems. The theory of finite-entanglement scaling dictates corrections to local observables that vanish algebraically in the bond dimension \cite{nishino1996numerical,tagliacozzo2008scaling,pollmann2009theory,pirvu,vanhecke2019scaling}. Finite-entanglement scaling with PEPS has recently also been explored for two-dimensional critical systems which are described by bosonic conformal field theory \cite{rader2018finite,corboz2018finite,czarnik2019finite,vanhecke2021scaling}. Here, we aim to investigate how efficiently we can approximate fermionic critical states, in particular those exhibiting Fermi surfaces.
	
	In fermionic systems, logarithmic violations of the area law of entanglement go hand in hand with the presence of codimension one Fermi surfaces \cite{wolfarealawviolation,klicharealawviolation}. 
	These discontinuities in the system's momentum distribution manifest themselves already in translation-invariant, quadratic models as continuous sets of zero energy modes in Fourier space. However, the relation between the presence of Fermi surfaces and the entanglement scaling is not qualitatively altered by the presence or absence of interactions. Therefore, we will focus on free-fermion systems with a Fermi surface, which allows for the application of the Gaussian and fermionic version of the PEPS \textit{Ansatz} (GfPEPS) \cite{gfpepskraus}, thereby reducing the computational cost of the required simulations. We first show that in one dimension these states can reproduce the aforementioned power-law improvement of the precision as a function of the bond dimension by considering the critical points of the Kitaev chain. Subsequently, both 1D and 0D Fermi surfaces in 2D lattice systems are treated by considering the $p$-wave superconductor. In both cases, we again obtain a power-law relation between bond dimension and precision in the thermodynamic limit (albeit with different exponents), indicating that PEPS can describe gapless models and in particular Fermi surfaces of arbitrary dimensions.
	
\begin{figure}
		\begin{center}
			\includegraphics[width=0.2\textwidth]{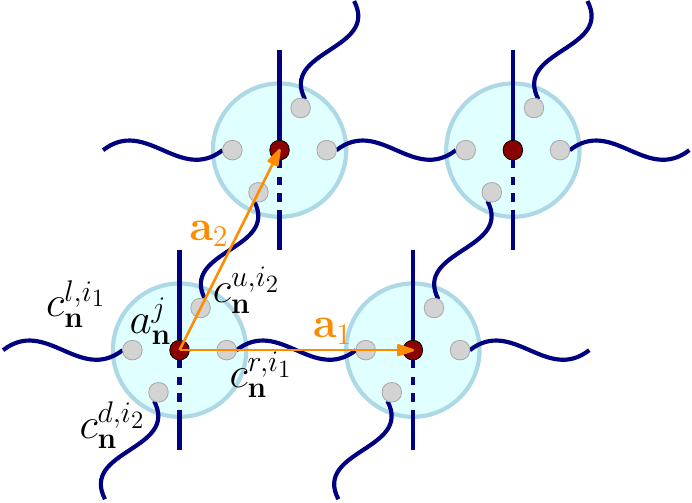}
			\caption{Schematic of a GfPEPS on a 2D lattice with unit vectors $\mathbf{a}_1$ and $\mathbf{a}_2$. Majorana modes (gray balls) are entangled (blue lines) to form a maximally correlated state which is locally projected by a Gaussian map (big blue circles) to the physical fermions (red balls).}
			\label{gfpeps_schem}
		\end{center}
	\end{figure}

	\paragraph{Gaussian fermionic PEPS} --- Consider a 2D lattice built up by a periodic repetition of $N_1\times N_2=N$ unit cells, spanned by $\mathbf{a}_1$ and $\mathbf{a}_2$. To each vertex we attribute $f$ physical fermionic orbitals with creation (annihilation) operators $a^{j^\dagger}_\mathbf{n}\,\left(a^{j}_\mathbf{n}\right)$ where $\mathbf{n}=n_i \mathbf{a}_i$ with $n_i=0,...,N_i-1$ is the site and $j=1,...,f$ the orbital index. Corresponding Majorana operators are denoted by $c^{2j-1}_\mathbf{n}= a^{j^\dagger}_\mathbf{n}+a^{j}_\mathbf{n}$ and $c^{2j}_\mathbf{n}=-i\left(a^{j^\dagger}_\mathbf{n}-a^{j}_\mathbf{n}\right)$. Within this framework, a PEPS \textit{Ansatz} is obtained by first introducing four sets of virtual Majoranas per site: $\{c^{l,{i_1}}_\mathbf{n}\}$, $\{c^{r,{i_1}}_\mathbf{n}\}$, $\{c^{d,{i_2}}_\mathbf{n}\}$ and $\{c^{u,{i_2}}_\mathbf{n}\}$ with $i_1=1,...,\chi_1$ and $i_2=1,...,\chi_2$. Next, a maximally correlated state $\rho_\text{in}$ is constructed on the virtual level by entangling neighboring Majoranas in both directions (see Fig.\,\ref{gfpeps_schem}). This is realized by placing the Majoranas in their joint vacuum, essentially creating $\chi_i$ virtual Majorana chains in the direction of $\mathbf{a}_i$. Finally, the maximally correlated state is locally projected onto the physical level by a channel $\mathcal{E}=\bigotimes_\mathbf{n}\mathcal{E}^\text{loc}_\mathbf{n}$ encoding the fermionic PEPS tensor and yielding the (possibly mixed) $\rho_\text{out}=\mathcal{E}\left(\rho_\text{in}\right)$ (see Supplemental Material \cite{suppmat} for more details on $\rho_\text{in}$ and $\mathcal{E}$). By increasing the number of virtual Majoranas, the variational set can be enlarged. Note that, as the number of Majoranas can be different in each direction, the resulting effective bond dimensions, $D_i=\sqrt{2}^{\chi_i}$, can differ as well.
	
	Since $\rho_\text{in}$ is a free-fermion state, Gaussianity of the PEPS can be enforced by restricting the channel $\mathcal{E}$ to be Gaussian as well \cite{bravyi,gaussianorig}. Not only can then both the input and the output state be fully described in terms of their real and antisymmetric correlation matrices, $\Gamma^{ij}_{\mathbf{n}\mathbf{m}}=\frac{i}{2} \text{Tr}\left(\rho \left[c^i_\mathbf{n},c^j_\mathbf{m}\right]\right)$, but there is also a link between both, prescribed by $\mathcal{E}$, in the form of a Schur complement, $\Gamma_\text{out}=A+B\left(D+\Gamma_\text{in}^{-1}\right)^{-1} B^T \,.$
	Here, $\mathcal{E}$ is in fact parametrized by $A=\bigoplus_\mathbf{n}A^\text{loc}_\mathbf{n}$ and analogous decompositions apply for $B$ and $D$ with $A^\text{loc}_\mathbf{n}\in\mathbbm{R}^{2f\times2f}$, $B^\text{loc}_\mathbf{n}\in\mathbbm{R}^{2f\times2(\chi_1+\chi_2)}$ and $D^\text{loc}_\mathbf{n}\in\mathbbm{R}^{2(\chi_1+\chi_2)\times2(\chi_1+\chi_2)}$. Furthermore, $X=\begin{pmatrix}
		A&B\\
		-B^T&D
	\end{pmatrix}$ is antisymmetric and $XX^T\leq\mathbbm{1}$ with the equality holding for a pure state.
	
	For translation-invariant Gaussian states, it is more convenient to work in Fourier space where these states can be described completely in terms of the Fourier transformed correlation matrix,  $G^{ij}_{\mathbf{k}\mathbf{q}}=\frac{i}{2}\text{Tr}\left(\rho\left[d^i_\mathbf{k},d^{j^\dagger}_\mathbf{q}\right]\right)$. Herein, $d^i_\mathbf{k}=\frac{1}{\sqrt{N}}\sum_\mathbf{n}e^{-i\mathbf{k}\cdot\mathbf{n}}c^i_\mathbf{n}$ with momentum modes $\mathbf{k}$. In the case of periodic boundary conditions these are given by $\mathbf{k}=\sum_i\frac{k_i}{N_i}\mathbf{b}_i$ where $\mathbf{b}_i$ are the reciprocal lattice vectors and $k_i=0,...,N_i-1$, whereas for antiperiodic boundary conditions $\mathbf{k}=\sum_i \frac{1}{N_i}(k_i+\frac{1}{2})\mathbf{b}_i$. The Fourier transformed correlation matrix is anti-Hermitian, $GG^\dagger\leq\mathbbm{1}$ (with the equality again holding for a pure state) and for translation-invariant states $G$ decomposes in diagonal blocks $G=\bigoplus_\mathbf{k}G(\mathbf{k})$ with, for instance,
	\begin{equation}
		G_\text{in}(\mathbf{k}) = \begin{pmatrix}
			0 & e^{i \mathbf{k} \cdot \mathbf{a}_1}\\
			-e^{-i \mathbf{k} \cdot \mathbf{a}_1} & 0
		\end{pmatrix}^{\oplus \chi_1} \oplus \begin{pmatrix}
			0 & e^{i \mathbf{k} \cdot \mathbf{a}_2}\\
			-e^{-i \mathbf{k} \cdot \mathbf{a}_2} & 0
		\end{pmatrix}^{\oplus \chi_2}
		\label{Gin}
	\end{equation}
	for the input state \cite{suppmat}. Assuming translation invariance of the PEPS, so that $\mathcal{E}^\text{loc}_\mathbf{n}$ is independent of $\mathbf{n}$, the transition matrix $X$ decomposes into identical blocks and yields $G_\text{out}(\mathbf{k})=A^\text{loc}+B^\text{loc}\left(D^\text{loc}-G_\text{in}(\mathbf{k})\right)^{-1}B^\text{loc}$ where the purity of the input state was used to replace $G^{-1}_\text{in}(\mathbf{k})$ by $-G_\text{in}(\mathbf{k})$.
	
	As real-space correlation matrices are real valued, their Fourier transformed analogues have the property that $G(-\mathbf{k})=G^\ast(\mathbf{k})$. This implies that $G(\mathbf{k})$ is real and antisymmetric in points where $\mathbf{k}=-\mathbf{k}$, the time-reversal invariant modes (TRIMs). For pure Gaussian states, these $G(\mathbf{k})$ can thus be interpreted as ordinary, pure correlation matrices with a definite parity $\langle P_\mathbf{k}\rangle=\langle(-1)^{\sum_j a^{j^\dagger}_\mathbf{k}a^{j}_\mathbf{k}}\rangle=\langle\prod_{j}i\,d^{j^{(1)}}_\mathbf{k} d^{j^{(2)}}_\mathbf{k}\rangle=\text{Pf}(G(\mathbf{k}))$. Any pure fermionic Gaussian state hence has a specific TRIM parity configuration. E.g.\:\:the $G_\text{in}(\mathbf{k})$ of the input state [Eq.\,(\ref{Gin})] amounts to $\langle P_\mathbf{k}\rangle_\text{in} = \text{Pf}(G_\text{in}(\mathbf{k}))=\left(e^{i \mathbf{k}\cdot\mathbf{a}_1}\right)^{\chi_1}\left(e^{i\mathbf{k}\cdot\mathbf{a}_2}\right)^{\chi_2}$ so that the center of the Brillouin zone always has an even parity, $\langle P_\mathbf{0}\rangle_\text{in}=1$, while for the other TRIMs $\bigl<P_{\frac{\mathbf{b}_i}{2}}\bigr>_\text{in}=(-1)^{\chi_i}$ and $\bigl<P_{\frac{\mathbf{b}_1+\mathbf{b}_2}{2}}\bigr>_\text{in}=(-1)^{\chi_1+\chi_2}$. A $\frac{\mathbf{b}_i}{2}$ jump in Fourier space thus corresponds to an extra factor $(-1)^{\chi_i}$. Remarkably, this virtual parity configuration is lifted to the physical level by the Gaussian channel $\mathcal{E}$. Indeed, in order for the full pure PEPS \textit{Ansatz} to have a fixed global parity, the projectors from the virtual to the physical level (i.e., the Kraus operators of  $\mathcal{E}^\text{loc}_\mathbf{n}$) are designed to be parity conserving (changing)~\cite{gfpepskraus}. Combining this with their local and translation-invariant nature, GfPEPS have the same (opposite) parity configuration as the valence bond state. We conclude that pure, regular GfPEPS can only realize $2^d\times2$ (even/odd $\chi_i$ in each direction and parity conserving or changing $\mathcal{E}$) parity configurations while for an arbitrary, pure Gaussian state, there are $2^{2^d}$ possible configurations. Certain parity configurations thus cannot be reached by GfPEPS in spatial dimensions larger than 1, unless singular behavior is present \cite{dubail,wahl1,wahl2,wahl3}.
	
\begin{figure}
		\begin{center}
			\includegraphics[width=0.49\textwidth]{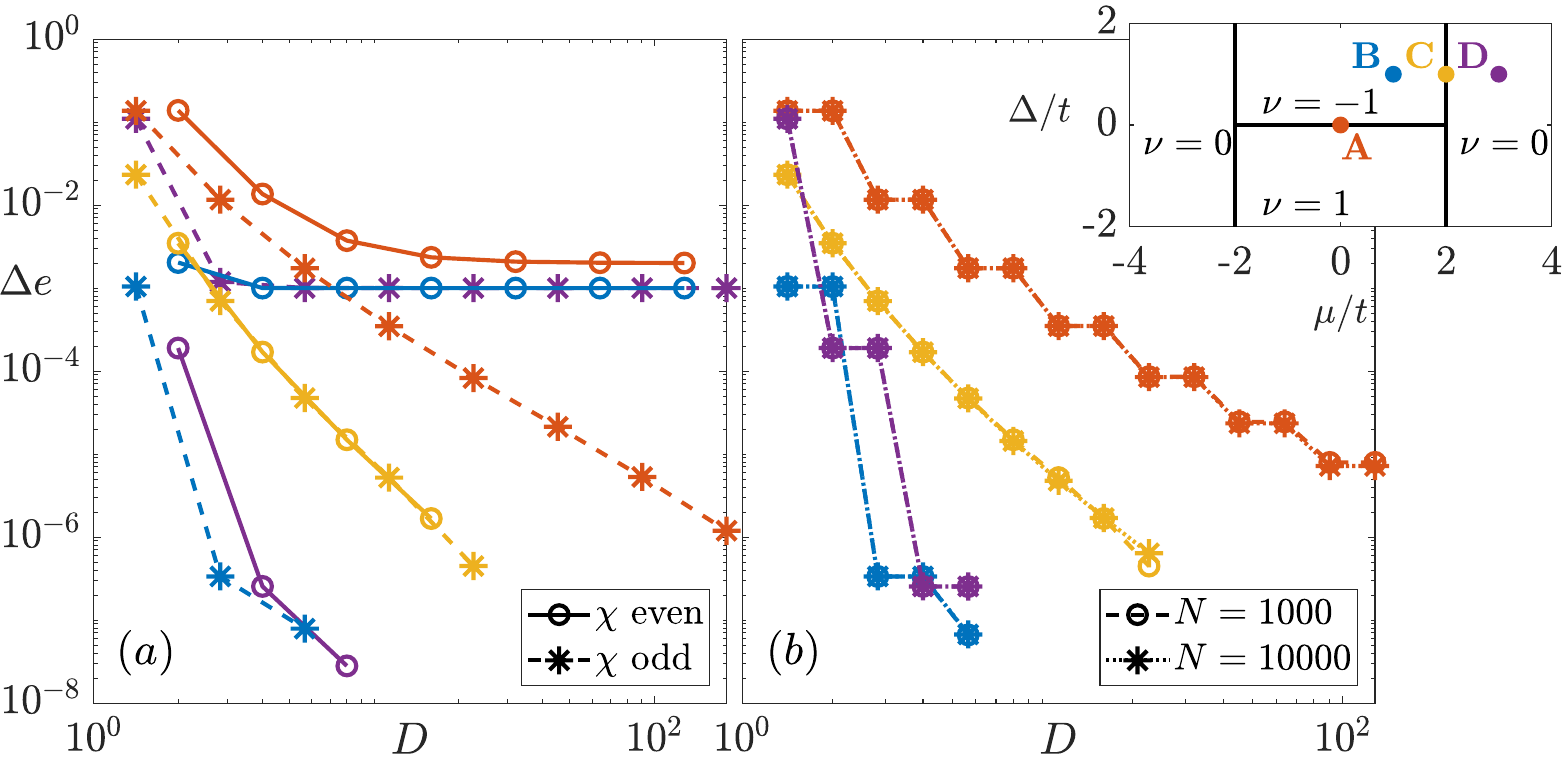}
			\caption{(a) Energy density error of optimized GfMPS for a Kitaev chain with $N=1000$ and periodic boundary conditions at 4 different points in the phase diagram (shown in the inset) as a function of the bond dimension $D=\sqrt{2}^\chi$. GfMPS with a $\chi$ that reproduces the exact parity configuration can approximate the exact solution with an arbitrary precision whereas the precision of parity obstructed GfMPS saturates to values of order $1/N$. (b) Energy density error of optimized GfMPS for a Kitaev chain with anti-periodic boundary conditions. As the TRIM in $k=0$ is avoided, fixed energy errors due to incorrect parity configurations do not occur. Results for two system sizes were compared, showing that the thermodynamic limit is well probed at $N=1000$.}
			\label{kitaev_conv}
		\end{center}
	\end{figure}

	\paragraph{Kitaev chain} --- Consider the 1D Kitaev chain of length $N$ with Hamiltonian $H = -t \sum_{n=0}^{N-1} \left(a_n^\dagger a_{n+1} +h.c.\right)   -\Delta \sum_{n=0}^{N-1} \left(a_n^\dagger a_{n+1}^\dagger + h.c.\right) -\mu \sum_{n=0}^{N-1} a_n^\dagger$. Further assume that $t$ is positive and that (anti-)periodic boundary conditions, $a_{n+N} = (-1)\, a_n$, apply. In Fourier space this can be expressed as $H = \sum_k \Upsilon^\dagger_k \left(\mathbf{h}(k) \cdot \boldsymbol{\sigma}\right) \Upsilon_k -\frac{\mu}{2}N$
	where $\mathbf{h}(k)=\left(-2t\cos k,-2i\Delta \sin k, -\mu \right)$, $\Upsilon_k = \left(a_k \quad a^\dagger_{-k}\right)^T$ and the $k$ values are as prescribed in the previous section. For $|\mu|>2t$, the system is in a trivial phase with the ground state reducing to a product state when $\Delta=0$ and all momentum modes filled (empty) when $\mu>2t$ ($\mu<-2t$). For $\Delta \neq 0$ and $|\mu|<2t$, on the other hand, the system is in a topological phase with winding number $|\nu| = 1$ and an isolated gapless Majorana mode on both ends of the wire when the chain is cut. Critical lines lie at $\mu = \pm 2t$ and at $\Delta=0$ when $|\mu|<2t$ (Fig.\,\ref{kitaev_conv}). We optimized GfMPS by minimizing their energy density for both gapless and gapped parameter choices and for a number of virtual Majorana modes, $\chi$, ranging from 1 to 15. The resulting energy density errors, defined as the difference between the GfMPS energy density and that of the exact ground state at the system size under consideration, are displayed in Fig.\,\ref{kitaev_conv}.
	
	In the left panel, periodic boundary conditions apply and for the critical hopping model at $(t,\mu,\Delta)=(1,0,0)$ (point A in the phase diagram), a power-law improvement of the precision is obtained in the case of an odd number of virtual Majoranas. An even $\chi$, on the other hand, yields a saturating profile for $\Delta e$. Similar observations apply in the chiral phase (point B) but with an even faster convergence as the area law of entanglement is not violated. In particular, GfMPS constitute an exact  ground state with $\chi=1$ when $\Delta = \pm t, \mu = 0$. For gapless Hamiltonians on the critical lines between the chiral and trivial regions [e.g.,point C at $(t,\mu,\Delta)=(1,2,1)$], we obtain a power-law improvement of the precision but with a higher exponent than for the critical line between the two topological phases. These findings exemplify well-known results about the approximation of one-dimensional critical points with MPS where the rate of convergence depends on the conformal field theory (and, in particular, the central charge) underlying the critical point. Note that point $C$ is indeed an Ising-like transition (with conformal charge $c=1/2$), whereas point $A$ corresponds to a massless Dirac fermion (with $c=1$). Finally, optimization in the trivial phase (point D) results in profiles similar to those in the chiral phase but with the odd and even $\chi$ curves interchanged.
	
	In the previous section we explained how the number of virtual Majorana modes determines the parity configuration of the regular GfMPS \textit{Ansatz}: an even (odd) $\chi$ should be used when the parity in the two TRIMs is equal (opposite). When using a $\chi$ with the wrong parity, one TRIM (and thus $1$ out of $N$ modes in momentum space) cannot have the correct correlation matrix, leading to an energy density error of order $1/N$, which is confirmed in Fig.\,\ref{kitaev_conv}(a). Only singular behavior with nonanalyticities in the problematic TRIMs can circumvent the fixed parity structure but this fine-tuned case is not supported by our variational method \cite{suppmat}. There is also a link between the parity configuration and the $\mathbbm{Z}_2$ invariant characterizing the topological features of the model \cite{kitaev_2001}.  Here, this is reflected by the fact that an even $\chi$ should be used in the trivial phase while an odd $\chi$ should be utilized in the topological regions, in accordance with the relevant underlying physics with isolated Majorana edge modes. When using antiperiodic boundary conditions, the TRIM in the zone center is never sampled. As a result, fixed energy errors due to an incorrect parity configuration will never occur and the resulting energy convergence curves will not saturate. This is confirmed in Fig.\,\ref{kitaev_conv}(b) and thus proves to be the most pragmatic solution to study the convergence of the energy precision. Again, we report a seemingly exponential improvement in the gapped models whereas a power-law scaling (with different exponents in A and C) is obtained at criticality. Finally, note that adding one extra virtual Majorana to a GfMPS with a correct parity configuration does not improve the energy precision (in points A, B and D). Indeed, this extra virtual Majorana chain decouples completely from the physical system, yielding a singular norm-zero state when closed with periodic boundary conditions. Closing with antiperiodic boundary conditions on the other hand yields a nonzero norm but with the same energy as with one Majorana less. Only in point C, which is on the critical line between the phases with opposing parity configurations, does the addition of one virtual Majorana improve the results. Indeed, as the TRIM at $k=\pi$ coincides with the Fermi surface at this critical line, its ground state parity is not uniquely defined.
	
\begin{figure}
		\begin{center}
			\includegraphics[width=0.49\textwidth]{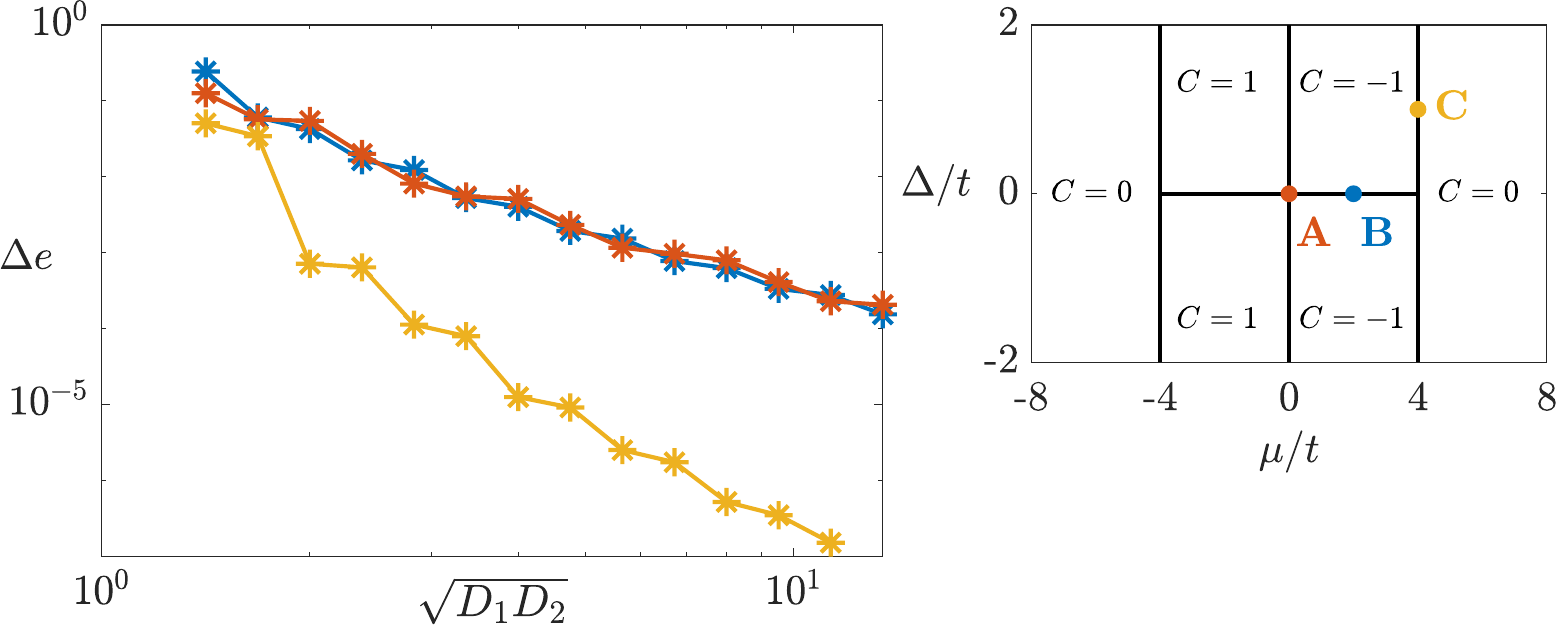}
			\caption{Energy density error of optimized GfPEPS for the $p$-wave superconductor (left) (with its phase diagram on the right panel) and linear size $L=1000$ (thus probing the thermodynamic limit) as a function of the geometric mean of the bond dimensions $\sqrt{D_1D_2}$. In all the considered cases the energy precision improves polynomially with $\sqrt{D_1D_2}$, albeit with different exponents for A and B with 1D Fermi surfaces and C with a Dirac cone.}
			\label{pwave_conv}
		\end{center}
	\end{figure}

	\paragraph{p-Wave superconductor} --- Switching to 2D, the analogue of the Kitaev chain is the $p$-wave superconductor on a square lattice. The Hamiltonian has an identical Fourier space description but now with $\mathbf{h}(\mathbf{k})=-2\left(t\left(\cos k_x+\cos k_y\right),i\Delta \left(\sin k_x+i\sin k_y\right), \frac{\mu}{2} \right)$ where the phase difference between the two spatial components of the pairing term is necessary to open a gap. Again taking $t>0$, topologically, trivial regions are found when $|\mu|>4t$. For $0<\mu<4t$, one obtains a chiral phase with Chern number $C=-1$, whereas $-4t<\mu<0$ yields $C=+1$. Critical lines lie in between and at analogous places as for the 1D Kitaev chain (Fig.\,\ref{pwave_conv}). For $\Delta=0$ and $-4t<\mu<4t$ the model exhibits a 1D Fermi surface. In order to circumvent fixed energy errors related to problematic parity configurations, we will only work with antiperiodic boundary conditions in both directions so that for any system size the TRIM in the Brillouin zone center is not sampled. Furthermore, the utilized linear system sizes are always even so that also the other TRIMS are avoided. 
	
	\begin{figure}
		\begin{center}
			\includegraphics[width=0.49\textwidth]{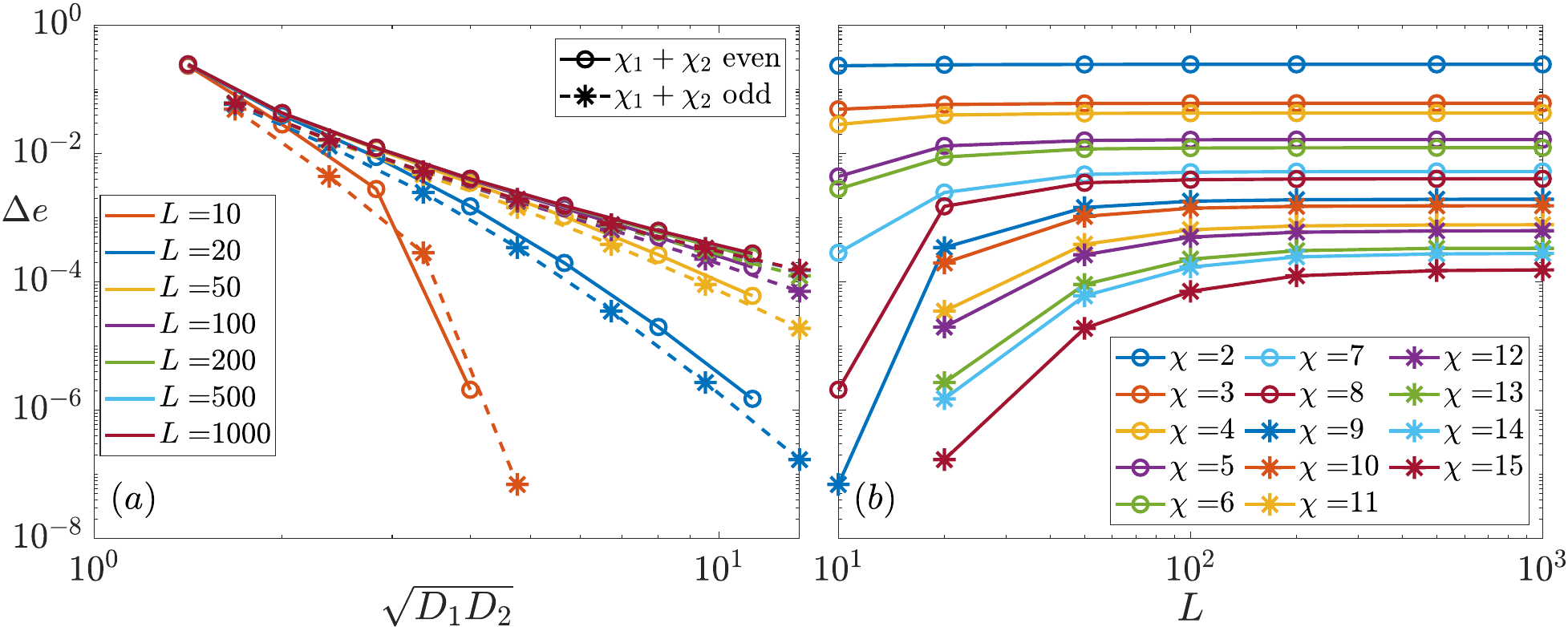}
			\caption{Energy density error of optimized GfPEPS for the square lattice hopping model (point A in Fig.\,\ref{pwave_conv}) as a function of (a) the geometric mean of the bond dimensions $\sqrt{D_1D_2}$ and (b) the linear system size $L$. In the latter, $\chi = \chi_1+\chi_2$ with $\chi_1=\chi_2 (+ \, 1)$ when $\chi$ is even (odd). The thermodynamic limit is probed for all bond dimensions when $L=1000$ and the onset of a power law can be discerned in the left panel. Results for even and odd $\chi$ are displayed separately to obtain smooth curves.}
			\label{pwave_convA}
		\end{center}
	\end{figure}
	
\begin{figure}
		\begin{center}
			\includegraphics[width=0.49\textwidth]{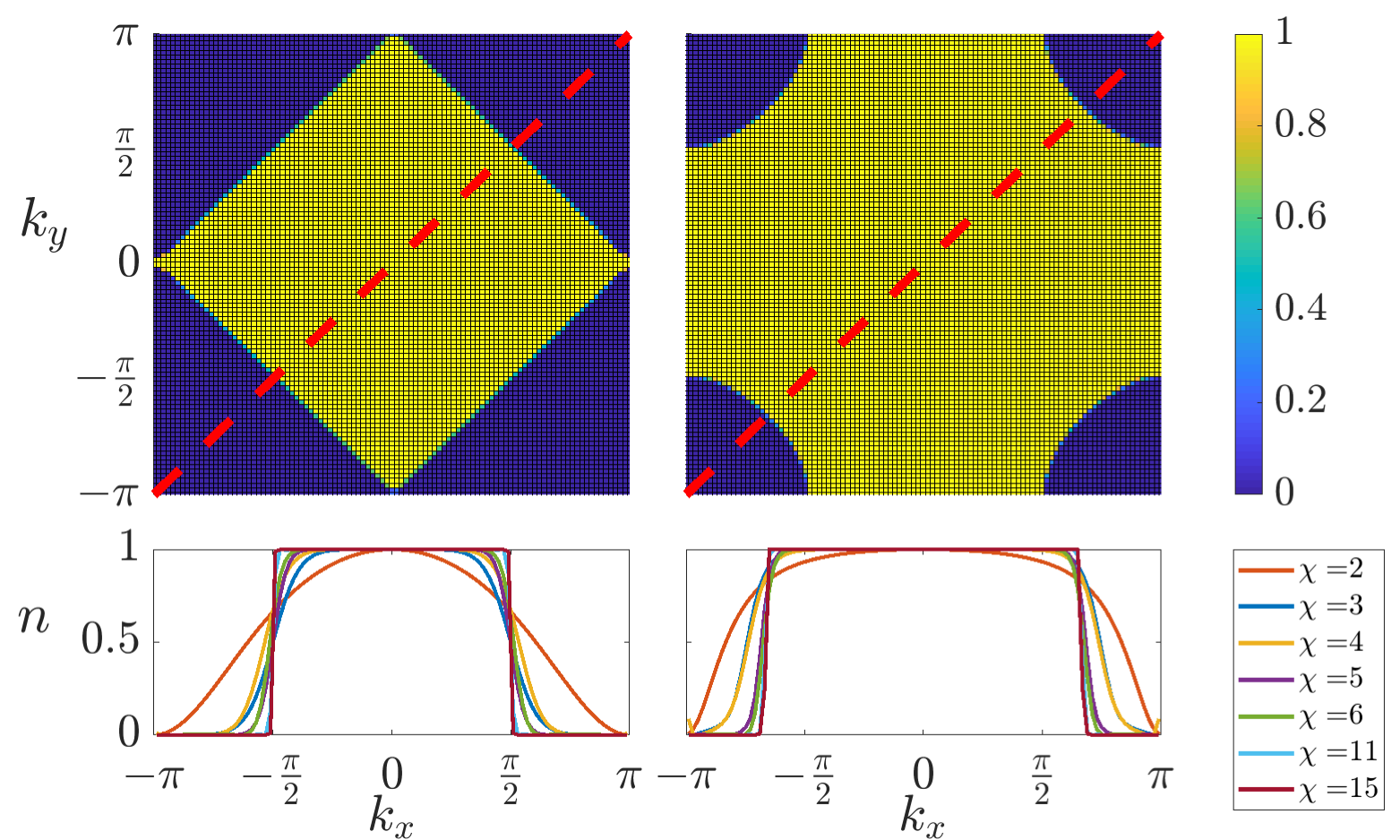}
			\caption{The top panels display the occupation number $n(\mathbf{k}) =  \frac{1}{2} \left[1 + \frac{i}{2}\left(G^{11}(\mathbf{k})-G^{22}(\mathbf{k})\right) + \frac{1}{2}\left(G^{12}(\mathbf{k})+G^{21}(\mathbf{k})\right)\right]$ of the optimized GfPEPS with the highest bond dimension, $(\chi_1,\chi_2)=(8,7)$, for the spinless, square lattice hopping model with commensurate (left) and incommensurate (right) filling (i.e., points A and B in Fig.\,\ref{pwave_conv}, respectively). In the bottom panels, occupations are compared for different combinations of Majorana numbers along the diagonal of the Brillouin zone (red line where $k_x=k_y$), showing that the sharp edges of the Fermi surface are reproduced increasingly well.}
			\label{FS_conv}
		\end{center}
	\end{figure}

	GfPEPS were first optimized for point A where $\mu=0$ and the model reduces to spinless fermions hopping on a square lattice, exhibiting a one-dimensional Fermi surface that divides the Brillouin zone in a filled and a vacated half. The energy density errors as a function of both the geometric mean of the bond dimensions, $\sqrt{D_1 D_2}$, and the linear system size, $L=\sqrt{N}=N_1=N_2$, are displayed in Fig.\,\ref{pwave_convA}. The right panel shows that by increasing the system size and keeping $\sqrt{D_1 D_2}$ fixed, the energy error saturates, indicating that the thermodynamic limit is probed. For the largest system size, this is the case for all the considered bond dimensions and the curve for $L=1000$ in Fig.\,\ref{pwave_convA}(a) can hence be taken as the energy density error as a function of $\sqrt{D_1 D_2}$ in the thermodynamic limit. Herein, a power-law improvement of the precision can clearly be discerned. This curve was also copied in Fig.\,\ref{pwave_conv} where we compare it to results obtained in a similar way for the B and C points in the phase diagram. The B point was studied because the exact parity configuration of the target state cannot be reproduced by GfPEPS in this case due to the incommensurate filling. The GfPEPS will thus approximate singular behavior. However, as we used antiperiodic boundary conditions, this does not spoil the energy convergence study and Fig.\,\ref{pwave_conv} confirms that the energy precision increases according to the same power law as for the square lattice hopping model. We conclude that 2D models with 1D (and thus codimension one) Fermi surfaces can (even in the thermodynamic limit) be approximated by PEPS with an arbitrary precision by increasing the bond dimension in a controlled way. To solidify this claim even more, Fig.\,\ref{FS_conv} displays the occupation number $\langle n(\mathbf{k})\rangle = \langle a_\mathbf{k}^\dagger a_\mathbf{k}\rangle$ of the GfPEPS with the highest bond dimension, $(\chi_1,\chi_2)=(8,7)$, for point A and B, clearly showing that the Fermi surfaces are resolved successfully. In the bottom panels the filling profile along the diagonal of the Brillouin zone is compared for multiple bond dimensions, again demonstrating that by increasing the bond dimension, the sharp edges of the Fermi surface are reproduced to a good degree (see also Supplemental Material\cite{suppmat} for additional results). Point C, on the other hand, is interesting because in this case the criticality exists only in one $\mathbf{k}$ mode, essentially realizing a 0D Fermi surface with a linear dispersion around it, i.e., a Dirac cone. Optimizing GfPEPS at point C shows that energy precision again increases according to a power law. Just as in the 1D case the exponent of this power law is higher than in the A and B points. Indeed, the area law of entanglement is not violated in C.
	
	\paragraph{Conclusions} --- We studied whether projected entangled-pair states can be used to describe critical systems exhibiting Fermi surfaces. This question was answered in an affirmative way. Indeed, the Gaussian and fermionic version of the PEPS \textit{Ansatz} was successfully applied to 2D free-fermion systems with both 1D and 0D Fermi surfaces. More specifically, we considered two critical points of the $p$-wave superconductor with a 1D Fermi surface and observed that in the thermodynamic limit the precision of the GfPEPS approximations increased according to similar power laws as a function of the bond dimension. This is the 2D extension of earlier results hereabout in 1D, that were also reproduced for the gapless Kitaev chain. Furthermore, 0D Fermi surfaces exhibiting Dirac cones were also shown to pose no difficulties for PEPS as the energy precision for yet another critical point of the $p$-wave superconductor also increased according to a power law but with an even higher exponent.
	
	Though we did not address the ability of fermionic PEPS to approximate interacting systems with a Fermi surface directly, the qualitative features of the convergence that we obtain should be robust against adding interactions. Indeed, it has been amply demonstrated that the success of PEPS (and tensor networks more generally) is not affected by the strength of interactions, but by the scaling of entanglement. As interactions are not expected to affect the entanglement scaling of critical points and Fermi surfaces \cite{swingle2010,ding2012}, neither should be the ability for PEPS to approximate them. Moreover, energy densities obtained in this work present an upper bound to those a generic (non-Gaussian) PEPS would be able to attain for the same (quadratic) Hamiltonian. Indeed, it can be expected that discarding the free-fermion structure would further improve the accuracy as already demonstrated for the 1D case \cite{francorubio2022}. Furthermore, the variationally optimized Gaussian PEPS obtained in this work can play a significant role also in the case of interacting systems. Indeed, the ground state of an interacting system can first be approximated in a mean-field-like manner by a Gaussian PEPS, which is then converted to a generic (fermionic) PEPS tensor \cite{gfpepskraus} (using e.g.\ the formalism of super vector spaces \cite{z2graded}), in order to serve as the initial state for a full-fledged variational optimization over the set of all PEPS. Note that the effective bond dimensions of the Gaussian PEPS in our simulations should be within reach of current state-of-the-art PEPS algorithms and make this a feasible approach, that will be investigated further in forthcoming work.
	
	\begin{acknowledgements}
		
		We would like to thank Erez Zohar, Karel Van Acoleyen and Ignacio Cirac for inspiring discussions. This work has received support from the European Research Council (ERC) under the European Union's
		Horizon 2020 program [Grant Agreements No.\ 715861 (ERQUAF), 647905 (QUTE), 636201 (WASCOSYS), and 863476 (SEQUAM)], and from the DFG
		(German Research Foundation) under Germany's Excellence Strategy (EXC2111--390814868).
		
	\end{acknowledgements}
	
	\bibliography{manuscript.bib}
	
\end{document}